\documentclass[journal,transmag]{IEEEtran}
\makeatletter
\newcommand*{\rom}[1]{\expandafter\@slowromancap\romannumeral #1@}
\makeatother
\newtheorem{theorem}{Theorem}
\newtheorem{definition}{Definition}

\newtheorem{lemma}{Lemma}
\newtheorem{remark}{Remark}
\newtheorem{example}{Example}
\usepackage{amsmath,amssymb}
\usepackage{cite}
\usepackage{tikz}
\usetikzlibrary{shapes,arrows}
\usepackage{verbatim}
\usepackage{caption}
\usepackage{pgfplots}
\usepackage{bbm}
\allowdisplaybreaks

\newcommand{\Hz}{\mathcal{H}_0}
\newcommand{\Ho}{\mathcal{H}_1}
\newcommand{\Hj}{\mathcal{H}_j}

\newcommand{\mh}[1]{{\color{blue}#1}}

\title{Testing Against Independence with An Eavesdropper}

\author{\IEEEauthorblockN{Sara Faour\IEEEauthorrefmark{*},  Mustapha Hamad\IEEEauthorrefmark{*}, 
Mireille Sarkiss\IEEEauthorrefmark{**}, and Mich\`ele Wigger\IEEEauthorrefmark{*}}
\IEEEauthorblockA{\IEEEauthorrefmark{*}LTCI, Telecom Paris, IP Paris, sara.faour@ip-paris.fr, mustapha.hamad7@gmail.com, michele.wigger@telecom-paris.fr}
\IEEEauthorblockA{\IEEEauthorrefmark{**}SAMOVAR, Telecom SudParis, IP Paris, mireille.sarkiss@telecom-sudparis.eu}
}

\begin{document}

\maketitle

\tikzstyle{circle} = [circle,minimum size=#1, inner sep=0pt, outer sep=0pt]
\tikzstyle{block} = [draw, rectangle, rounded corners, minimum height=2em, minimum width=5em]
\tikzstyle{vert_block} = [draw, rectangle,rounded corners, minimum height=5.5em, minimum width=5em]
\tikzstyle{input} = [coordinate]
\tikzstyle{output} = [coordinate]
\tikzstyle{texte} = [rectangle]
\tikzstyle{pinstyle} = [pin edge={to-,thin,black}]

\begin{abstract}
We study a distributed binary hypothesis testing (HT) problem with communication and security constraints, involving three parties: a remote sensor called Alice, a legitimate decision center called Bob, and an eavesdropper called Eve, all having their own source observations. In this system, Alice  conveys a rate-$R$ description of her observations to Bob, and Bob performs a binary hypothesis test on the joint distribution underlying  his and Alice's observations.  The goal of Alice and Bob is to 
 maximize the exponential decay of Bob's miss-detection (type-II error)  probability under two constraints: Bob's false-alarm (type-I error) probability has to stay below a given threshold and Eve's uncertainty  (equivocation) about  Alice's observations should stay above a given security threshold even when Eve learns Alice's message.  For the special case of testing against independence, we characterize the largest possible type-II error exponent under the described type-I error probability and security constraints. 
\end{abstract}


\section{Introduction}

\IEEEPARstart{I}n future ultra-massive type communications, billions of IoT devices and sensors will be connected and cooperate together to detect, measure, and monitor environmental phenomena and events in distributed monitoring and alert systems.
The different events can be considered as different hypotheses and are assumed to determine the joint probability distribution underlying the data observed at the various nodes. We focus on binary  hypothesis testing  where we have two possible events: a normal situation,  called \textit{null hypothesis}  $\Hz$, and an alert situation, called \textit{alternative hypothesis} $\Ho$. In this case, there are two types of errors. Type-\rom{1} error refers to the event that the decision center decides on $\Hz$  while the true hypothesis is $\Ho$. Type-\rom{2} error refers to the event that the decision center  decides on $\Ho$ while the true hypothesis is $\Hz$. 

We consider in this paper \emph{distributed hypothesis testing (DHT)} with a single sensor Alice and a single decision center Bob, each observing an independently and identically distributed (i.i.d.)  source sequence, where  the two sequences are  jointly drawn according to the known  probability mass function  $P_{XY}$ under hypothesis $\Hz$ and according to the product of the marginals $P_X P_Y$ under $\Ho$.  Information-theorists refer to this setup as \emph{testing against independence}. Alice can send a rate-$R$ message to Bob describing her observations and aiming to help Bob in deciding on the true hypothesis. The focus here is on the Stein exponent, i.e.,
 on the largest possible exponential decay for Bob's type-II error probability under the requirement that his type-I error probability stays below a given threshold $\epsilon \in (0,1)$. This largest possible type-II error exponent in this setup was determined by Ahlswede and Csisz\'ar \cite{Ahlswede} and does not depend on the value of $\epsilon$. In this paper, we consider an extension of the Ahlswede-Csisz\'ar result to a setup including an additional eavesdropper Eve that observes a local i.i.d. source sequence, intercepts Alice's message to Bob $M$, and wishes to learn about Alice's source sequence $X^n$. In this extended setup, Alice is required to choose her message in a way that Eve's equivocation about the source $X^n$ stays above pre-determined thresholds given the two hypothesis. 

Hypothesis testing has also been considered under other security constraints. In particular, the works in  \cite{Sreekumar_Privacy, Sankar, Tandon, Selma, Oechtering, Vinod} focused  on ensuring data privacy in various forms. For instance, \cite{Selma} considered a model where a sensor has to pre-randomize its data before using it on the distributed hypothesis testing problem. In \cite{Sreekumar_Privacy}, not the sensor's data but only a related information has to be kept private from the decision center, either in an average distortion or equivocation sense. The work in \cite{Vinod} allowed for interactive communication and applied a privacy constraint inspired by the cryptography literature. 

The secrecy scenario with an external eavesdropper that we study in the present paper, was already treated in \cite{piantanida} and in \cite{TACI_HT} for the more general scenario of testing against conditional independence. As we show, in the special case of testing against independence, the type-II error exponents proposed in \cite{piantanida, TACI_HT} are optimal in the limit of vanishing type-I error probabilities $\epsilon \to 0$ but are generally suboptimal for fixed $\epsilon>0$.  For general $\epsilon>0$,  the optimal exponent is achieved by using the scheme in \cite{piantanida, TACI_HT} with probability $(1-\epsilon)$ and using a degenerate scheme with probability $\epsilon$. In this degenerate scheme,  Alice  sends a dummy zero-message\mh{,} and upon receiving this message\mh{,} Bob declares the alternative hypothesis $\Ho$. The converse is shown through a change-of-measure argument and by proving asymptotic Markov chains, similar to the converse proofs in \cite{Mustapha, strong_converse}, see also \cite{Tyagi}. In this paper, we however need extra non-trivial steps for the converse  bounds on the equivocation  under the two hypotheses.


\textit{Notation:}
We follow standard notations. In particular, we  denote by $o(1)$  any function that tends to 0 as $n\to \infty$. Also, we denote by $\mathcal{T}^{(n)}_{\mu}(P_{XY})$ the strongly typical set defined in  \cite{Ahlswede}, and we abbreviate  $\mathcal{T}^{(n)}_{n^{-1/3}}(P_{XY})$ simply by $\mathcal{T}^{(n)}(P_{XY})$.  We further abbreviate \emph{probability mass function} by \emph{pmf}. When the pmf is not clear from the context, we write $H_P(\cdot)$ and $I_P(\cdot;\cdot)$ to indicate that entropy and mutual information are meant with respect to $P_{XY}$.

\section{Problem Setup and Main Result}

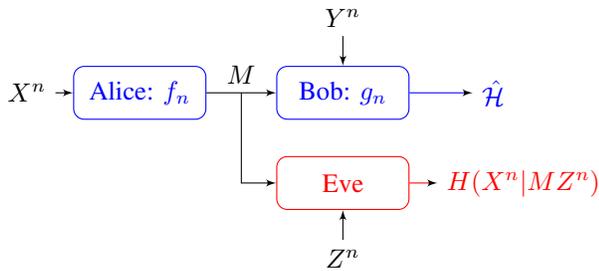
\begin{figure}[!t]
\centering
\begin{tikzpicture}[auto, node distance=2cm,>=latex']
	\node [texte](textA) {$X^n$};
    \node [block, right of=textA,draw=blue, text=blue, node distance=1.5cm] (encoder) {Alice: $f_n$};
    \node [block, right of= encoder, draw=blue, text=blue, node distance=2.7cm] (bob) {Bob: $g_n$};
    \node [block, below of=bob, draw=red, text= red, node distance=1.2cm] (eve) {Eve};
    \node [texte,above of= bob,node distance=1cm](textY) {$Y^n$};
    \node [texte, below of= eve,node distance=1cm](textE) {$Z^n$};
    \node [texte,right of= bob, text=blue](h) {$\hat{\mathcal{H}}$};
    \node [texte, right of= eve,text=red, node distance =2.4cm](entro) {$H(X^n|MZ^n)$};
    
    \draw [->] (textA) -- (encoder);
    \draw [->] (textY) -- (bob);
    \draw [->] (textE) -- (eve);
    \draw [->] (encoder) -- node[name=b] {$M$} (bob);
    \draw [->] (b) |- (eve);
    \draw [->,draw=blue] (bob) -- (h);
    \draw [->,draw=red] (eve) -- (entro);
\end{tikzpicture}
\captionsetup{justification=centering}
\caption{DHT with communication and security constraints.} \label{fig:setup}
\end{figure}
Consider the  DHT setup illustrated in Figure~\ref{fig:setup} involving the three terminals Alice, Bob, and Eve. Depending on the binary hypothesis $\Hz$ or $\Ho$, the observations at the three terminals obey the following joint distribution 
\begin{IEEEeqnarray}{rCl}\textnormal{under } \Hz \colon \quad \left(X^{n}, Y^{n}, Z^n\right)& \sim&  P_{X Y Z}^{\otimes n}\\
\textnormal{under } \Ho\colon \quad \left(X^{n}, Y^{n}, Z^n\right)& \sim& Q_{XYZ}^{\otimes n},\end{IEEEeqnarray}
where $Q_{XYZ}=P_{X}P_Y P_{Z|XY}$.

 Alice observes the independent and identically distributed (i.i.d.) length-$n$ sequence $X^{n}$ and sends $M = f_{n}({X}^{n})$ for some randomized encoding function of the form $f_n: \mathcal{X}^{n} \rightarrow \mathcal{M}$ and message space $\mathcal{M} \triangleq\{1,\ldots, \lceil 2^{n R}\rceil \}$, where $R>0$ is the maximum allowed rate of transmission. Given its own observation $Y^{n}$ and after observing message $M$, Bob guesses  the true hypothesis as $\hat{\mathcal{H}}= g_{n}\left(M, Y^{n}\right)$ using a decision rule of the form $ \mathcal{M} \times \mathcal{Y}^{n} \rightarrow \{ \Hz,\Ho\}$. Bob's type-I and type-II error probabilities are then given by
\begin{IEEEeqnarray}{rCl}
 \alpha_{n}\left(f_{n}, g_{n}\right)&:=&\mathbb{P}(\hat{\mathcal{H}}=\Ho | \Hz) \\
\beta_{n}\left(f_{n}, g_{n}\right)&:=&\mathbb{P}(\hat{\mathcal{H}}=\Hz| \Ho).
\end{IEEEeqnarray}


\begin{definition}
\label{def 1}
Given $\epsilon>0$, a tuple $\left(R, \theta, \Delta_{0}, \Delta_{1}\right)$ is achievable, if there exists a sequence of encoding and decoding functions $\{(f_{n},g_{n})\}_n$  satisfying
\begin{subequations}\label{eq:constraints}
\begin{IEEEeqnarray}{rCl}
	 \varlimsup _{n \rightarrow \infty} \alpha_{n}(f_n,g_n) & \leq & \epsilon\\
 \varliminf _{n \rightarrow \infty} - \frac{1}{n}\log {\beta}_{n} (f_n,g_n) & \geq &\theta \label{1} \\
 \varliminf _{n \rightarrow \infty} \frac{1}{n}  H\left(X^{n} | M, Z^{n}, \Hj\right) &\geq & \Delta_j, \qquad j\in\{0,1\}.
\end{IEEEeqnarray}
\end{subequations}

\end{definition}

\begin{theorem}
\label{prop1}
For $\epsilon \in(0,1)$, the quadruple $(R, \theta, \Delta_0, \Delta_1)$ is achievable if, and only if, there exists a conditional pmf $P_{U|X}$ so that 
\begin{IEEEeqnarray}{rCl}
R &\geq & I_P(U ; X), \label{136}\\
\theta &\leq & I_P(U ; Y), \label{137} \\
\Delta_{0} & \leq &(1-\epsilon)H_{P}(X|UZ) +\epsilon H_{P} \left(X|Z \right), \label{138} \\
\Delta_{1} & \leq & (1-\epsilon)H_{Q}(X|UZ) +\epsilon H_{Q} \left(X|Z \right),\label{139}
\end{IEEEeqnarray}
where indices $P$ and $Q$ refer to the joint pmfs 
\begin{IEEEeqnarray}{rCl}
	P_{UXYZ} & = & P_{U|X}P_{XYZ}\\
	Q_{UXYZ} &=&P_{U|X}P_{X}P_{Y}P_{Z|XY}.
	\end{IEEEeqnarray}
\end{theorem}

\begin{remark}
In the limit $\epsilon\to 0$ and for $\Delta_1\geq H_{Q}(X|Z)$, the fundamental rate-exponent-equivocations region in Theorem~\ref{prop1} recovers the regions presented in 
 \cite{piantanida,TACI_HT}, which only considered an equivocation constraint under the null hypothesis $\Hz$.  For a  general positive $\epsilon>0$\mh{,} the   fundamental rate-exponent-equivocations region in Theorem~\ref{prop1} however is larger, unless  $\Delta_0$  is sufficiently small.
\end{remark}

\begin{figure}[!t]
	\centering
	\resizebox{!}{5.5cm}{
		\input{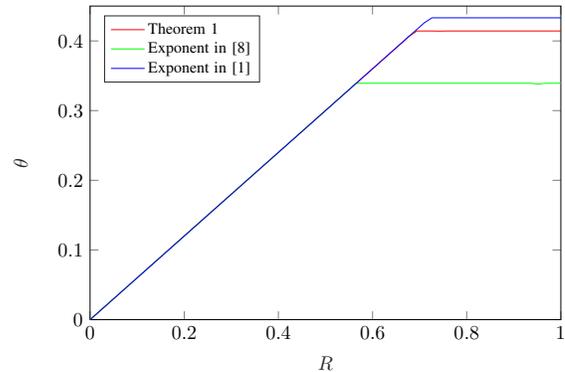}}
	\captionsetup{justification=centering}
	\caption{Type-II error exponent $\theta$ in function of $R$.} \label{fig:plot0}
\end{figure}

We evaluate Theorem~\ref{prop1} for a specific example. 
\begin{example}
Consider  a binary source $X$, and assume that $Y$ and $Z$ are obtained by passing $X$ through   a binary erasure channel (BEC) and a binary symmetric channel (BSC), respectively. Source and  channel parameters are given by
\begin{IEEEeqnarray}{rCl}
P_X(0) =1-P_X(1)&=0.8,\\
P_{Y|X}(e|x)& = 0.4,\\
P_{Z|X}(1-x|x)& =0.2,\\
Q_{Z|X}(1-x|x)& =0.3.
\end{IEEEeqnarray}
We also fixed  $\Delta_0=\Delta_1=0.13$ and $\epsilon=0.2$. For this example, the equivocation constraint under $\Ho$ is always less stringent than under $\Hz$.

Figure \ref{fig:plot0} shows the largest exponent $\theta$ for which the quadruple $(R, \theta, \Delta_0, \Delta_1)$ is achievable according to Theorem~\ref{prop1}, and compares it to the exponent proposed in \cite{piantanida} and the largest exponent achievable without  any security constraints \cite{Ahlswede}. For small rates $R$, all three exponents coincide and the equivocation  constraints under both hypotheses seem inactive.  For larger rates $R$\mh{,} the optimal exponent in Theorem~\ref{prop1} dominates the sub-optimal exponent in \cite{piantanida} because $\epsilon>0$. For even larger rates $R$, the security constraints become stringent the exponent in Theorem~\ref{prop1} is below the Ahlswede-Csisz\`ar exponent in \cite{Ahlswede}.
\end{example}

\section{Optimal Coding Scheme}

Choose a conditional pmf $P_{U|X}$ so that
\begin{IEEEeqnarray}{rCl}
R >I_P(U;X)
\end{IEEEeqnarray}
where we defined the joint pmf 
\begin{IEEEeqnarray}{rCl}
P_{UXYZ}& =& P_{U|X}P_{XYZ}.	
\end{IEEEeqnarray}

\textbf{Codebook generation:} Independently generate $\lceil 2^{nR}\rceil$ sequences $u^n(1),\ldots, u^n( \lceil 2^{nR}\rceil)$ by picking each entry of each sequence i.i.d. according to $P_U$. 
Denote the realization of the set of codewords $\mathcal{C}$\mh{.}

\textbf{Encoder Alice:} Fix a small value $\mu >0$.  Alice behaves in a randomized way, described by a Bernoulli-($1-\epsilon$) random variable $\Xi$  and the likelihood  encoder corresponding to the  chosen codebook $\mathcal{C}$ \cite{likelihood,Schieler_SRD}
\begin{IEEEeqnarray}{rCl}
P_{M'|X^n}^{\textnormal{LE},\mathcal{C}}(m|x^n)&= &\frac{P_{X|U}^{\otimes n}(x^n|u^n(m))}{\sum_{{m} \in\{1,\ldots,  \lceil 2^{nR} \rceil\}} P_{X|U}^{\otimes n}(x^n|u^n(m))} .
\end{IEEEeqnarray}

If $\Xi=0$\mh{,}  then 
Alice sends $M=0$. 
Otherwise,  for $X^n=x^n$\mh{,} it picks $M'\in\{1,\ldots,  \lceil 2^{nR} \rceil\}$ according to the conditional distribution $P_{M'|X^n}^{\textnormal{LE},\mathcal{C}}(\cdot|x^n)$. If the pair $(u^n(M'),x^n) \in \mathcal{T}^{(n)}(P_{UX})$, then Alice sends $M=M'$ and otherwise she sends $M=0$.
%

\textbf{Decoder Bob:} Assume $Y^n=y^n$ and $M=m$. Bob declares $\hat{H}=\Hz$ if $m\neq 0$ and $(u^n(m), y^n ) \in \mathcal{T}^{(n)}_{2\mu}(P_{UY})$. Else it declares $\hat{\mathcal{H}}=\Ho$. 

\textbf{Sketch of Analysis:} Given $\Xi=0$, the analysis is simple. Trivially, the type-II error probability equals 0 and the type-II error probability equals 1. Moreover, equivocations under the two hypotheses are $H_P(X|Z)$ and $H_Q(X|Z)$. 
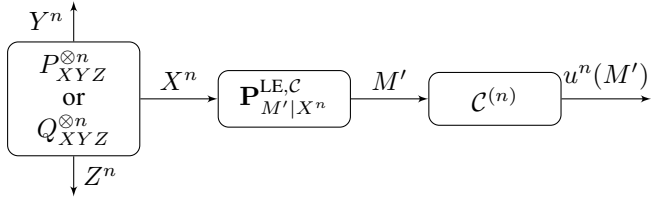
\begin{figure}[h]
\vspace{-1mm}
\centering
\begin{tikzpicture}[auto, node distance=1cm,>=latex']
	\node[block] (distribution) { $\begin{matrix}P_{XYZ}^{\otimes n}\\ \textnormal{or} \\Q_{XYZ}^{\otimes n}\end{matrix}$};
    \node [block,right of=distribution, node distance=2.8cm ] (encoder) {$\mathbf{P}^{\textnormal{LE},\mathcal{C}}_{M'|X^n}$};
        \draw [->] (distribution) -- node[name=(X)] {$X^n$} (encoder);
            \draw[->] (distribution) --  node[name=(Y)] {$Y^n$} (0,1.3);
                        \draw[->] (distribution) --  node[name=(Z)] {$Z^n$} (0,-1.3);

            \node [block, right of=encoder, node distance=2.8cm ] (Cu) {$\mathcal{C}^{(n)}$};
        \draw [->] (encoder) -- node[name=(M')] {$M'$} (Cu);
    \node [ right of=Cu, node distance=2.2cm] (end) {};
    \draw [->] (Cu) -- node[name=(u)] {$u^n(M')$} (end);
%
%
%
\end{tikzpicture}
\vspace{-6mm}

\captionsetup{justification=centering}
\caption{Encoding } \label{fig:real}
\end{figure}

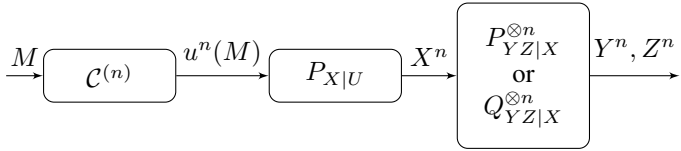
\begin{figure}[h]
\centering
\begin{tikzpicture}[auto, node distance=1.4cm,>=latex']
    \node  (input) {};
    \node [block, right of=input, node distance=1.5cm] (codebook) {$\mathcal{C}^{(n)}$};
    \draw [->] (input) -- node[name=M] {$M$} (codebook);
    \node [block, right of=codebook, node distance=3cm] (p1) {$P_{X|U}$};
    \draw [->] (codebook) -- node[name=v] {$u^n(M)$} (p1);
    \node [vert_block, right of=p1, node distance=2.5cm] (p2) {$\begin{matrix}P_{YZ|X}^{\otimes n}\\ \textnormal{or} \\ Q_{YZ|X}^{\otimes n}\end{matrix}$};
    \draw [->] (p1) -- node[name=x] {$X^n$} (p2);
    
    \node [ right of=p2, node distance=2.2cm] (end) {};
    \draw [->] (p2) -- node[name=(YZ)] {$\begin{matrix} Y^n, Z^n \end{matrix}$} (end);

\end{tikzpicture}
\captionsetup{justification=centering}
\caption{Idealized distribution.} \label{fig:ideal}
\end{figure}

Given $\Xi=1$, the analysis is similar to \cite{Sreekumar_Privacy} and based on the soft covering lemma in \cite{Schieler_SRD}. The likelihood encoding system is depicted in Figure~\ref{fig:real}.  Since $R>I(U;X)$, the pair  $(u^n(M'),X^n)$ is jointly typical under both hypotheses with a probability (when averaged over the random code construction) tending exponentially fast to 1 as $n \to \infty$. One can thus restrict the analysis to this assumption.  Moreover, since $R>I(U;X)$, by the generalized  soft-covering lemma in \cite{Schieler_SRD},   on average over the random code construction  
the joint pmf induced  by the real system in Figure~\ref{fig:real} is close to the pmf induced by the idealized system in Figure~\ref{fig:ideal}. 


By standard arguments, it can be concluded that on the idealized system and when $u^n(M)\in\mathcal{T}^{(n)}(P_{U})$, then the type-II error probability exponent is equal to $\theta = I(U;Y)$. The type-I error probability tends to 0 as $n\to \infty$ simply by the weak law of large numbers. Equivocation on the \emph{idealized system} under $\Hz$ is bounded as follows: 
\begin{IEEEeqnarray}{rCl}
\frac{1}{n}H(X^n|U^n(M)Z^n)& =&\frac{1}{n} \sum_{i=1}^n H(X_i |u_i(M)Z_i) \\
&= &H_P(X|UZ) +o(1),
\end{IEEEeqnarray}
where the first equality holds by the memorylesness of the channels and the second equality because $P_{U_i(M)}$ tends to $P_U$ as $n\to \infty$ as mentioned above.  Combining all these observations concludes the proof.
\section{Converse Proof to Theorem 1}

Fix an achievable exponent $\theta<\theta_{\epsilon}^{*}(R)$ and a sequence of (random) encoding and decision functions so that \eqref{eq:constraints} are satisfied. Further fix a blocklength $n>0$ and let $M$ and $\hat{\mathcal{H}}$ be the message and the guess produced by the chosen encoding and decision functions for this given blocklength. 

Define the set 
\begin{IEEEeqnarray}{rCl}
\mathcal{D}_{n}  \triangleq  \left\{ \left(x^{n}, y^n\right) \in \mathcal{T}^{(n)}_{n^{-1/3}} (P_{XY}) \colon g_n(f_n(x^n),y^n) =\Hz \right\}\IEEEeqnarraynumspace .\label{eq:Dn}
\end{IEEEeqnarray}
By the constraint on the type-I error probability and since 
by \cite[Lemma~2.12]{csiszar_korner_2011}
\begin{IEEEeqnarray}{rCl}
P_{XY}^{\otimes n}\left(\mathcal{T}_{n^{-1/3}}^{(n)}(P_{XY})\right) \geq 1-\frac{|\mathcal{X}||\mathcal{Y}|}{4 n^{1/3}},
\end{IEEEeqnarray}
we obtain  by the basic laws of probability
\begin{IEEEeqnarray}{rCl}
\Lambda_n := P_{XY}^{\otimes n}\left(\mathcal{D}_{n}\right) \geq 1-\epsilon-\frac{|\mathcal{X}||\mathcal{Y}|}{4 n^{1/3}}. \label{102}
\end{IEEEeqnarray}

Let  $( \tilde{X}^{n}, \tilde{Y}^{n})$ be the restriction of the pair $\left(X^{n}, Y^{n}\right)$ to $\mathcal{D}_{n}$,  $\tilde{M}=f_n(\tilde{X}^n)$ the new message, and  $\tilde{Z}^n$ the output of the discrete memoryless channel (DMC) $P_{Z|XY}$ for input sequences $(\tilde{X}^n, \tilde{Y}^n)$. 
Under $\Hz$, the probability distribution of the  quadruple $(\tilde{M}, \tilde{X}^{n}, \tilde{Y}^{n},\tilde{Z}^n)$ is  
\begin{IEEEeqnarray}{rCl}
\lefteqn{P_{\tilde{M} \tilde{X}^{n} \tilde{Y}^{n}\tilde{Z}^n}\left(m, x^{n}, y^{n},z^n\right) \triangleq} \quad\nonumber \\
&& P_{X Y}^{\otimes n}\left(x^{n}, y^{n}\right) \cdot \frac{\mathbbm{1}\left\{\left(x^{n}, y^n\right) \in \mathcal{D}_{n}\right\}}{\Lambda_n} \Pr[f_n(x^n)=m]. \IEEEeqnarraynumspace\label{eq:distr}
\end{IEEEeqnarray}
Let $T$ be uniform over $\{1,\ldots, n\}$ independent of all other random variables. 
\begin{lemma}\label{lem1}
For the distribution in \eqref{eq:distr}, the following limits hold as $n\to \infty$:
	\begin{IEEEeqnarray}{rCl}
		P_{\tilde{X}_T\tilde{Y}_T} &\to & P_{XY} \label{eq:pr3}\\
	\left|	\frac{1}{n}  H(\tilde{X}^n \tilde{Y}^n)-  H(\tilde{X}_T\tilde{Y}_T)\right| &\to& 0 \label{eq:pr1} \\
		 \left|	\frac{1}{n}  H(\tilde{Y}^n)-  H(\tilde{Y}_T)\right| &\to& 0 \label{eq:pr1b} \\
\left	|	\frac{1}{n} H(\tilde{X}^n|\tilde{Y}^n) - H(\tilde{X}_T|\tilde{Y}_T)\right| &\to& 0. \label{eq:pr2}	
		\end{IEEEeqnarray}
\end{lemma}
\begin{IEEEproof} See Appendix~\ref{app:Proof_lemma}.
			\end{IEEEproof}
	We bound the rate, the type-II error exponent and the equivocation based on Lemma~\ref{lem1}. 
	
	\textbf{Rate:} Throughout the following paragraphs, all quantities are calculated according to the pmf in \eqref{eq:distr} or the pmf $P_{XYZ}$, and we shall not mention this explicitly. For the rate we have:
\begin{IEEEeqnarray}{rCl}
	R & \geq&\frac{1}{n} H(\tilde{M}) = \frac{1}{n}I(\tilde{M}; \tilde{X}^n\tilde{Y}^n)\\
	& =& \frac{1}{n} H(\tilde{X}^n \tilde{Y}^n) -  \frac{1}{n} H(\tilde{X}^n\tilde{Y}^n|\tilde{M}) \\
	& = & H(\tilde{X}_T \tilde{Y}_T) +  o(1)  - \frac{1}{n} \sum_{t=1}^n  H( \tilde{X}_t \tilde{Y}_t| \tilde{X}^{t-1} \tilde{Y}^{t-1}\tilde{M})\IEEEeqnarraynumspace\\
	& =& H(\tilde{X}_T \tilde{Y}_T)+  o(1) - H(\tilde{X}_T \tilde{Y}_T| \tilde{X}^{T-1} \tilde{Y}^{T-1}\tilde{M}T)\IEEEeqnarraynumspace\\[1ex]
	&= & I(\tilde{X}_T \tilde{Y}_T;  \tilde{X}^{T-1} \tilde{Y}^{T-1}\tilde MT) +o(1)\\
	&\geq & I(\tilde{X}_T ;U)+o(1),\label{eq:rr}
\end{IEEEeqnarray}
where we defined $U \triangleq  (\tilde{X}^{T-1},\tilde{Y}^{T-1},\tilde M,T)$.
		
 To bound the error exponent, define $\tilde{\mathcal{H}}\triangleq g_n(\tilde{M}, \tilde{Y}^n)$ and notice  inequality 
 \begin{IEEEeqnarray}{rCl}
			D ( P_{\tilde{Y}^n \tilde{M}} \| P_{\tilde{Y}^n} P_{\tilde{M}} ) &\stackrel{(a)}{\geq} &D ( P_{\tilde{Y}^n \tilde{M}} (\tilde{\mathcal{H}}) \|  P_{\tilde{Y}^n} P_{\tilde{M}} ( \tilde{\mathcal{H}}) ) \IEEEeqnarraynumspace  \\
			& \stackrel{(b)}{=}   & 1\cdot  \log   \frac{1}{P_{\tilde{Y}^n} P_{\tilde{M}} ( \tilde{\mathcal{H}}=0) }, \label{eq:int}
		\end{IEEEeqnarray}
		where $(a)$ holds by the data-processing inequality and $(b)$ holds  by the definition of divergence and because   $\tilde{\mathcal{H}}=0$ with probability 1.

\textbf{Type-II error exponent:} We have:
		 \begin{IEEEeqnarray}{rCl}
		\beta_n & = & - \frac{1}{n} \log   P_{{Y}^n} P_{{M}} ( \hat{\mathcal{H}}=0)    \\
			&\stackrel{(c)}{\leq }  & - \frac{1}{n} \log   P_{\tilde{Y}^n} P_{\tilde{M}} ( \tilde{\mathcal{H}}=0)  -  \frac{2}{n}  \log \Lambda_n \\
			& \stackrel{(d)}{\leq } &\frac{1}{n} D ( P_{\tilde{Y}^n \tilde{M}} \| P_{\tilde{Y}^n} P_{\tilde{M}} ) +o(1) \\
			& =& \frac{1}{n}I( \tilde{M}; \tilde{Y}^n)  +o(1)\\
					& \leq & \frac{1}{n} H(\tilde{Y}^n) - \frac{1}{n} \sum_{t=1}^n H(  \tilde{Y}_t|\tilde{M } \tilde{X}^{t-1} \tilde{Y}^{t-1} )+o(1)\IEEEeqnarraynumspace\\
			&\stackrel{(e)}{ \leq} & H(\tilde{Y}_T)+o(1) - H(\tilde{Y}_T |U) \\
			&=
	& I(\tilde{Y}_T ; U)+o(1) ,\label{eq:ee}
		\end{IEEEeqnarray}
where $(c)$ holds because 
\begin{IEEEeqnarray}{rCl}
P_{\tilde{Y}^n} (y) \leq \frac{P_{Y^n}(y^n)}{\Lambda_n} \quad \textnormal{and} \quad 
	P_{\tilde{M}} (m) \leq \frac{P_{M}(m)}{\Lambda_n}; 
	\end{IEEEeqnarray}
	 $(d)$ holds by \eqref{102} and \eqref{eq:int};  and $(e)$ holds by \eqref{eq:pr1b}.
	
\textbf{Equivocation under $\Hz$:} We define 
$E\triangleq \mathbbm{1}\{(X^n,Y^n)\in \mathcal{D}_n\}$ and note:
\begin{IEEEeqnarray}{rCl}\label{Eq:XMZ}
\lefteqn{	\frac{1}{n}	H(X^n|M Z^n)}\nonumber\\
	&\stackrel{(f)}{=}	&\frac{1}{n}\sum_{t=1}^n H(X_t|X^{t-1}Y^{t-1}M Z^n) \\
	& = & \frac{1}{n}	\sum_{t=1}^n  H({X}_t|{X}^{t-1}{Y}^{t-1}M{Z}^nE)  \nonumber \\
	&& +  \frac{1}{n}	\sum_{t=1}^n  I(E;{X}_t|{X}^{t-1}{Y}^{t-1}M {Z}^n)\\
		& \le & \frac{1}{n}	\sum_{t=1}^n  H({X}_t|{X}^{t-1}{Y}^{t-1}M {Z}^nE)  \nonumber \\
		&& +  \frac{1}{n}	\sum_{t=1}^n  I(E;{X}_t,Y_t|{X}^{t-1}{Y}^{t-1}M {Z}^n)\\
			&\stackrel{(g)}{\leq} & \frac{1}{n}	\sum_{t=1}^n  H(\tilde{X}_t|\tilde{X}^{t-1}\tilde{Y}^{t-1}\tilde{M}\tilde {Z}^n)\Pr[E=1]   \nonumber \\
		&&+	 \frac{1}{n}	\sum_{t=1}^n  H({X}_t| {Z}_t,E=0)\Pr[E=0] \nonumber \\
		&& +  \frac{1}{n} I(E;{X}^nY^n|M{Z}^n)\\
	&\leq &H(\tilde{X}_T|U \tilde{Z}_T)\Pr[E=1]  \nonumber \\
	&& +H({X}_T| {Z}_T ,E=0)\Pr[E=0]    +\frac{1}{n} \label{eq:XMZ}
\end{IEEEeqnarray}
where $(f)$ holds by the Markov chain $X_t \to (M, X^{t-1}, Z^n) \to Y^{t-1}$; and $(g)$  because event $E=1$ corresponds to the change of measure in \eqref{eq:pr2} and because conditioning can only reduce entropy.

Define $F=1$ as the indicator function
\begin{IEEEeqnarray}{rCl}
F=\mathbbm{1}\{ (X^n,Z^n)\in \mathcal{T}^{(n)}(P_{XZ}) \}. 
\end{IEEEeqnarray}
Similarly to the proof of \eqref{eq:pr3}, one can show that 
\begin{IEEEeqnarray}{rCl}
P_{{X}_T{Z}_T| E=0,F=1} \to P_{XZ},
	\end{IEEEeqnarray}
	and thus by continuity of the entropy functional 
	\begin{IEEEeqnarray}{rCl}
	H(\tilde{X}_T|\tilde{Z}_T, E=0,F=1) \to H(X|Z). 
	\end{IEEEeqnarray}
	Since $H(\tilde{X}_T| \tilde{Z}_T ,E=0,F=0)$ is bounded by $\log|\mathcal{X}|$ and 
	\begin{IEEEeqnarray}{rCl}
	\Pr[ F=0,E=0] \leq \Pr[F=0] =o(1),
	\end{IEEEeqnarray}
	we conclude that 
	\begin{IEEEeqnarray}{rCl}
H(\tilde{X}_T| \tilde{Z}_T ,E=0)\Pr[E=0] \leq H(X|Z) \Pr[E=0] + o(1), \nonumber \\
	\end{IEEEeqnarray}
	which combined with \eqref{eq:XMZ} yields
	\begin{IEEEeqnarray}{rCl}
	{1\over n}H(X^n|MZ^n) & \leq & H(\tilde{X}_T|U  \tilde{Z}_T)\Pr[E=1]  \nonumber \\
	&& +H(X|Z) \Pr[E=0] +o(1).\label{eq:sum}
	\end{IEEEeqnarray}
	For sufficiently large values of the blocklength $n$, the conditional entropy $H(\tilde{X}_T|U  \tilde{Z}_T)$ is smaller than $H(X|Z)$ because $P_{\tilde{X}_T\tilde{Z}_T}\to P_{XZ}$, and thus \eqref{102} and \eqref{eq:sum} yields:
	\begin{IEEEeqnarray}{rCl}
			{1 \over n}	H(X^n|MZ^n) & \leq & H(\tilde{X}_T|U  \tilde{Z}_T)\left(1-\epsilon - \frac{|\mathcal{X}||\mathcal{Y}|}{4n^{{1/3}}}\right)\nonumber \\
				&& +H(X|Z) \left(\epsilon + \frac{|\mathcal{X}||\mathcal{Y}|}{4n^{{1/3}}} \right) +o(1).  \label{eq:equiv0}
	\end{IEEEeqnarray}

	\allowdisplaybreaks[4]
	\textbf{Equivocation under $\Ho$:}
	The proof is similar as under $\Hz$, but requires adding new random variables ${Y'}^n=(Y_1',\ldots, Y_n')$ obtained by passing $X^n$ through the DMC $P_{Y|X}$. We restrict the tuples $(X^n,{Y'}^n, Z^n,M)$ to tuples so that $(X^n,{Y'}^n) \in \mathcal{D}_n$ as introduced in  \eqref{eq:Dn}. Then the joint pmf under $\Ho$ of the  restricted tuple $(\bar X^n,\bar{Y'}^n, \bar Z^n,\bar M)$ is 
	\begin{IEEEeqnarray}{rCl}
\lefteqn{Q_{\bar{M} \bar{X}^{n} {\bar{Y'}}^{n}\bar{Z}^n}\left(m, x^{n},{y'}^n,z^n\right) } \quad\nonumber \\
&\triangleq& P_{X Y}^{\otimes n}\left(x^{n}, {y'}^n\right)\cdot \frac{\mathbbm{1}\left\{\left(x^{n}, {y'}^n\right) \in \mathcal{D}_{n}\right\}}{\Lambda_n}\nonumber \\
&& \cdot  Q_{Z|X}^{\otimes n}(z^n|x^n) P_{M|X^n}(m|x^n),\label{eq:distrQ}
\end{IEEEeqnarray}
where $Q_{Z|X}(z|x)=\sum_{y} P_{Y}(y) P_{Z|XY}(z|x,y)$.
	
	Following  the same steps as leading to \eqref{eq:equiv0},  but where$P_{XZ}$  is replaced by $Q_{XZ}=P_XQ_{Z|X}$, the sequence $Y^n$ by ${Y'}^n$, and  the restricted tuple $( \tilde{M}, \tilde{X}^{n} ,{\tilde{Y'}}^{n},\tilde{Z}^n)$ by $(\bar{M}, \bar{X}^{n} ,{\bar{Y'}}^{n},\bar{Z}^n)$, 
	we obtain an equivocation  bound under $\Ho$:
		\begin{IEEEeqnarray}{rCl}
			{1\over n}H_{Q}(X^n|MZ^n) & \leq & H(\bar{X}_T|\bar{U}  \bar{Z}_T)\left(1-\epsilon - \frac{|\mathcal{X}||\mathcal{Y}|}{4n^{1/3}} \right)\nonumber \\
			&& +H_Q(X|Z) \left(\epsilon + \frac{|\mathcal{X}||\mathcal{Y}|}{4n^{1/3}})\right)+o(1). \label{eq:equiv1} \IEEEeqnarraynumspace
		\end{IEEEeqnarray}
	Note that $(\bar{U}, \bar{X})$ have same  pmf as $(U, \tilde{X})$ defined previously, and $\bar{Z}_T$ is obtained by passing $\bar{X}$ through the DMC $Q_{Z|X}$.

	\textbf{Concluding the proof:}
	Before being able to  conclude the proof,  we notice the following set of inequalities (where again all pmfs are with respect to the pmf in \eqref{eq:distr}):
		\begin{IEEEeqnarray}{rCl}
			0 & = &\frac{1}{n} I(\tilde{M}; \tilde{Y}^n |\tilde{X}^n)   \\
			&= &\frac{1}{n} H( \tilde{Y}^n |\tilde{X}^n)  - \frac{1}{n} H( \tilde{Y}^n |\tilde{X}^n \tilde{M})  \\
			&= & H(\tilde{Y}_T|\tilde{X}_T) +o(1) -\frac{1}{n} \sum_{t=1}^n H( \tilde{Y}_t |\tilde{X}^n\tilde{Y}^{t-1} \tilde{M})\\
			& \geq& H(\tilde{Y}_T|\tilde{X}_T) +o(1) - H(\tilde{Y}_T | \tilde{X}_T \tilde{X}^{T-1}\tilde{Y}^{T-1}\tilde{M}T) \IEEEeqnarraynumspace\\
			& =&I(\tilde{Y}_T; U | \tilde{X}_T)  + o(1).  
		\end{IEEEeqnarray}
Thus, 
\begin{IEEEeqnarray}{rCl}\label{eq:mm}
	\lim_{n\to \infty} I(\tilde{Y}_T; U | \tilde{X}_T) =0.
	\end{IEEEeqnarray}
The proof is then concluded by combining  \eqref{eq:rr}, \eqref{eq:ee}, \eqref{eq:equiv0}, and \eqref{eq:equiv1} with limit \eqref{eq:mm} and taking $n\to \infty$. Details are as follows. By Carath\'eodory's theorem, and because $P_{\tilde{X}_T\tilde{U}}=P_{\bar{X}_T\bar{U}}$, we can conclude that  the existence of  a random variable $U_{n}$ over an alphabet of size $|\mathcal{X}|+3$ and so that 
\begin{IEEEeqnarray}{rCl}
	R &\geq &I_P({U}_n; \tilde{X}_{T}) +o(1)\\  
	-\frac{1}{n} \log \beta_{n} & \leq  & I_P({U}_n ; \tilde{Y}_{T}) +o(1) \\
 \varliminf _{n \rightarrow \infty}  H_P\left(X^{n} \mid M, Z^{n}\right)&\leq &H(\tilde{X}_T \mid {U}_n, \tilde{Z}_T)\\
 \varliminf _{n \rightarrow \infty}  H_Q\left(X^{n} \mid M, Z^{n}\right) & \leq & H(\tilde{X}_T \mid  U_n, \bar{Z}'_T),
\end{IEEEeqnarray} 
where  $\bar{Z}'_T$ is obtained by passing $\tilde{X}_T$ through the DMC $Q_{Z|X}$. 

Considering a subsequence of blocklengths $\{n_i\}_{i=1}^\infty$ for which the joint pmf  $P_{\tilde{X}_T\tilde{Y}_TU_n}$ converges, we conclude the existence of joint pmfs $P_{XYZU}$ and $Q_{XYUZ}$ with the properties desired in Theorem~\ref{prop1}. 
This concludes the proof of the converse.

\section{Conclusion}

We have studied the problem of distributed hypothesis testing  against independence over a rate-limited noiseless channel with both communication and security constraints. We have characterized the largest possible  type-II error exponent at the legitimate receiver under constraints on the legitimate receiver's type-I error probability   and the equivocations measured at an eavesdropper. In the limit of vanishing type-I error probability the results recover the previous result in \cite{piantanida}. This previous result is however disproved when positive type-I error probabilities are allowed.  

An interesting future research direction is to extend our results  to a scenario with variable-length coding, when the expected rate but not the maximum rate is constrained.
\appendices
	\section{Proof of Lemma~\ref{lem1}}\label{app:Proof_lemma}
	To prove \eqref{eq:pr3}, notice that 
	\begin{IEEEeqnarray}{rCl}
		P_{\tilde{X}_T\tilde{Y}_T}(x,y) & = & \frac{1}{n} \sum_{t=1}^n P_{\tilde{X}_t\tilde{Y}_t}(x,y) \\
		&= &\mathbb{E}\left[ \frac{1}{n} \sum_{t=1}^n \mathbbm{1}\{\tilde{X}_t =x , \tilde{Y}_t=y\} \right] \\
		&=& \mathbb{E}[\pi_{\tilde{X}^n\tilde{Y}^n}(x,y)].
	\end{IEEEeqnarray}
	Since by the definition of the typical set,  
	\begin{IEEEeqnarray}{rCl}
	|\pi_{\tilde{X}^n\tilde{Y}^n}(x,y) -P_{XY}(x,y)| \leq n^{-1/3},
	\end{IEEEeqnarray}
	we conclude that as $n\to \infty$ the probability $	P_{\tilde{X}_T\tilde{Y}_T}(x,y)$ tends to $P_{XY}(x,y)$. 
	
	To prove \eqref{eq:pr1}, notice first that 
	\begin{IEEEeqnarray}{rCl}
		\lefteqn{\frac{1}{n}H(\tilde{X}^n \tilde{Y}^n)+\frac{1}{n} D( P_{\tilde{X}^n \tilde{Y}^n} \| P_{XY}^{\otimes n} )} \nonumber \\
		& =& 
		- \frac{1}{n} \sum_{(x^n, y^n)\in \mathcal{D}_n} P_{\tilde{X}^n\tilde{Y}^n} (x^n,y^n) \log P_{XY}^{\otimes n} (x^n,y^n)\\
		&=& - \frac{1}{n}  \sum_{t=1}^n \sum_{(x^n, y^n)\in \mathcal{D}_n} P_{\tilde{X}^n\tilde{Y}^n} (x^n,y^n) \log P_{XY}(x_t,y_t) \IEEEeqnarraynumspace\\
		&=& - \frac{1}{n} \sum_{t=1}^n\sum_{ (x,y)\in\mathcal{X}\times \mathcal{Y}} P_{\tilde{X}_t\tilde{Y}_t} (x,y) \log P_{XY}(x,y) \\
		&=&  - \sum_{ (x,y)\in\mathcal{X}\times \mathcal{Y}} P_{\tilde{X}_T\tilde{Y}_T}(x,y) 
		\log P_{XY}(x,y)\\
		& =&  H(\tilde{X}_T\tilde{Y}_T) + D(P_{\tilde{X}_T\tilde{Y}_T} \| P_{XY}).
		\label{eq:fsum}		\end{IEEEeqnarray}
	Combined with the following two limits \eqref{eq:limit1} and \eqref{eq:limit2}, this establishes \eqref{eq:pr1}. The first relevant limit is 
	\begin{IEEEeqnarray}{rCl}
		D(P_{\tilde{X}_T\tilde{Y}_T}\| P_{XY}) \to 0,\label{eq:limit1}
	\end{IEEEeqnarray}
	which holds by \eqref{eq:pr3} and because $P_{\tilde{X}_T\tilde{Y}_T}(x,y)=0$ whenever $P_{XY}(x,y)=0$. 
	The second limit is:
	\begin{IEEEeqnarray}{rCl}
	\label{eq:limit2}
\frac{1}{n} D( P_{\tilde{X}^n \tilde{Y}^n} \| P_{XY}^{\otimes n})\to 0, \end{IEEEeqnarray}
	and holds because  $\frac{1}{n}\log \Lambda_n \to 0$ and by the following set of inequalities:  
	\begin{IEEEeqnarray}{rCl}
		0& \leq& \frac{1}{n} D( P_{\tilde{X}^n \tilde{Y}^n} \| P_{XY}^{\otimes n})  \nonumber \\
		&= &  \frac{1}{n} \sum_{(x^n, y^n)\in \mathcal{D}_n} P_{\tilde{X}^n\tilde{Y}^n} (x^n,y^n) \log \frac{P_{\tilde{X}^n\tilde{Y}^n} (x^n,y^n)}{P_{XY}^{\otimes n}(x^n,y^n)} \IEEEeqnarraynumspace \\
		& = &- \frac{1}{n} \sum_{(x^n, y^n)\in \mathcal{D}_n} P_{\tilde{X}^n\tilde{Y}^n} (x^n,y^n) \log \Lambda_n \\ 
		&=& -\frac{1}{n} \log \Lambda_n.\label{eq:limit2x}
	\end{IEEEeqnarray}

	To prove \eqref{eq:pr1b}, notice that by the same arguments as we concluded \eqref{eq:fsum}, we also have
	\begin{IEEEeqnarray}{rcl}	\frac{1}{n}H( \tilde{Y}^n)+\frac{1}{n} D( P_{\tilde{Y}^n} \| P_{Y}^{\otimes n} )  =H(\tilde{Y}_T) + D(P_{\tilde{Y}_T} \| P_{Y}).\IEEEeqnarraynumspace\label{eq:sum2}
	\end{IEEEeqnarray}
	Moreover, \eqref{eq:limit1} and \eqref{eq:limit2} imply
	\begin{IEEEeqnarray}{rCl}
		\frac{1}{n} D( P_{\tilde{Y}^n}\|P_{Y}^{\otimes n}) & \to & 0\\
		D(P_{\tilde{Y}_T} \| P_{Y}) & \to & 0,
	\end{IEEEeqnarray}
	which combined with \eqref{eq:sum2} imply \eqref{eq:pr1b}.
	
		The last limit \eqref{eq:pr2} follows by the chain rule and limits \eqref{eq:pr1} and \eqref{eq:pr1b}.
	This concludes the proof.
\clearpage
\bibliographystyle{IEEEtran}

\end{document}